\pdfoutput=1

\documentclass[11pt,twoside,a4paper,cmspaper,final,collab]{cms-tdr}
\def\svnVersion{374702}\def\cmsCernNoTag{CERN-EP/2016-005}\def\cmsCernDate{\today}\def\cmsMessage{Published in Physics Letters B as \href{http://dx.doi.org/10.1016/j.physletb.2016.06.039}{\doi{10.1016/j.physletb.2016.06.039.}}}

\begin{document}\cmsNoteHeader{EXO-14-013}

\hyphenation{had-ron-i-za-tion}
\hyphenation{cal-or-i-me-ter}
\hyphenation{de-vices}
\RCS$HeadURL: svn+ssh://svn.cern.ch/reps/tdr2/papers/EXO-14-013/trunk/EXO-14-013.tex $
\RCS$Id: EXO-14-013.tex 374612 2016-11-22 20:25:15Z skhalil $

\newlength\cmsFigWidth
\ifthenelse{\boolean{cms@external}}{\setlength\cmsFigWidth{0.98\columnwidth}}{\setlength\cmsFigWidth{0.65\textwidth}}
\ifthenelse{\boolean{cms@external}}{\providecommand{\cmsLeft}{top\xspace}}{\providecommand{\cmsLeft}{left\xspace}}
\ifthenelse{\boolean{cms@external}}{\providecommand{\cmsRight}{bottom\xspace}}{\providecommand{\cmsRight}{right\xspace}}
\ifthenelse{\boolean{cms@external}}{\providecommand{\cmsTableResize[1]}{\resizebox{\columnwidth}{!}{#1}}}{\providecommand{\cmsTableResize[1]}{\relax{#1}}}
\newcommand{\st}{\ensuremath{S_{\mathrm{T}}}\xspace}
\newcommand{\mumu}{\ensuremath{\PGm^{\pm}\PGm^{\mp}}\xspace}
\newcommand{\ee}{\ensuremath{\Pe^{\pm}\Pe^{\mp}\xspace}}
\newcommand{\emu}{\ensuremath{\Pe^{\pm}\PGm^{\mp}}\xspace}
\newcommand{\mstop}{\ensuremath{M_{\PSQt}}\xspace}
\newcommand{\mchi}{\ensuremath{M_{\PSGcpmDo}}\xspace}
\newcommand{\dm}{\ensuremath{\Delta M_{\PSQt, \PSGcpmDo}}\xspace}
\newcommand{\mll}{\ensuremath{M_{\ell\ell}}\xspace}
\newcommand{\stmin}{\ensuremath{S_{\mathrm{T}}^{\text{min}}}\xspace}
\newcommand{\nj}{\ensuremath{N_{\text{jets}}}\xspace}
\newcommand{\ETcone}{\ensuremath{E_{\mathrm{T},\text{cone}}}\xspace}

\ifthenelse{\boolean{cms@external}}{\providecommand{\suppMaterial}{Fig. 1 of the supplementary material [URL will be inserted by publisher]}}{\providecommand{\suppMaterial}{Fig.~\ref{fig:stmin} of Appendix~\ref{app:suppMat}}}

\cmsNoteHeader{EXO-14-013}

\title{Search for R-parity violating decays of a top squark in proton-proton collisions at $\sqrt{s} = 8$\TeV}

\date{\today}

\abstract{The results of a search for a supersymmetric partner of the top quark (top squark), pair-produced in
proton-proton collisions at $\sqrt{s} = 8$\TeV, are presented. The search, which focuses on R-parity violating,
chargino-mediated decays of the top squark, is performed in final states with low missing transverse momentum,
two oppositely charged electrons or muons, and at least five jets. The analysis uses a data sample corresponding
to an integrated luminosity of 19.7\fbinv collected with the CMS detector at the LHC in 2012. The data are found
to be in agreement with the standard model expectation, and upper limits are placed on the top squark pair production
cross section at 95\% confidence level. Assuming a 100\% branching fraction for the top squark decay chain,
$\PSQt\to \PQb\PSGcpmDo, \PSGcpmDo\to \ell^\pm+\mathrm{jj}$,
top squark masses less than 890 (1000)\GeV for the electron (muon) channel are excluded for the first time in models
with a single nonzero R-parity violating coupling $\lambda^{\prime}_{ijk}$ $(i,j,k \leq 2)$, where $i,j,k$ correspond to
the three generations.
}

\hypersetup{%
pdfauthor={CMS Collaboration},%
pdftitle={Search for R-parity violating decays of a top squark in proton-proton collisions at sqrt(s) = 8 TeV},%
pdfsubject={CMS},%
pdfkeywords={CMS, physics, supersymmetry, leptons, low missing transverse energy}}

\maketitle

\section{Introduction}
\label{sec:intro}

Supersymmetry (SUSY)~\cite{susy, Martin:1997ns} is an extension of the standard model (SM)
that may provide a solution to the hierarchy problem~\cite{Dimopoulos:1995mi,Carlos:1993yy}.
In the SUSY framework, quadratically divergent radiative corrections to the Higgs boson mass,
dominated by loops involving the top quark, are canceled by loops with a supersymmetric partner
of the top quark (top squark).
The mass of the top squark is expected to be
within a few hundred \GeV of the top quark mass, and the supersymmetric Higgs boson partners are also expected
to have masses less than 1\TeV~\cite{Papucci:2011wy, Kitano:2006gv}.

Searches for SUSY are performed in many decay channels and are classified into R-parity conserving (RPC) and R-parity
violating (RPV) scenarios. The quantum number, R-parity, $P_\mathrm{R} = (-1)^{3B+L+2s}$ has a value $+1$ for SM particles
and $-1$ for superpartners, where $B$, $L$, and $s$ are baryon number, lepton number, and spin, respectively~\cite{rpv}.
In RPC models the top squark is expected to decay into the lightest SUSY particle, which escapes
detection. This results in an event signature with substantial missing transverse momentum.
Recent searches performed at the LHC at CERN in events with high missing transverse momentum have reduced the
parameter space available for a low mass top squark
~\cite{Chatrchyan:2013mya,Khachatryan:2014doa,Chatrchyan:2013xna, Aad:2014bva, Aad:2014qaa, Aad:2014kra}.
However, R-parity may not be conserved, in which case searches for SUSY particle decaying to SM particles without substantial missing transverse
momentum are important.

The superpotential terms that result in R-parity violation are given by:
\ifthenelse{\boolean{cms@external}}{
\begin{multline}
\label{eqn:wrpv}
W_\mathrm{RPV} =
\frac{1}{2}\, \lambda_{ijk}L_iL_j\overline{E}_k+
\lambda^\prime_{ijk}L_iQ_j\overline{D}_k\\+
\frac{1}{2}\, \lambda^{\prime \prime}_{ijk}\overline{U}_i\overline{D}_j\overline{D}_k + \mu_i L_i H_{u};
\end{multline}
}{
\begin{equation}
\label{eqn:wrpv}
W_\mathrm{RPV} =
\frac{1}{2}\, \lambda_{ijk}L_iL_j\overline{E}_k+
\lambda^\prime_{ijk}L_iQ_j\overline{D}_k\,+\,
\frac{1}{2}\, \lambda^{\prime \prime}_{ijk}\overline{U}_i\overline{D}_j\overline{D}_k + \mu_i L_i H_{u};
\end{equation}
}
where $\Lam_{ijk}$, $\Lamp_{ijk}$, and $\Lampp_{ijk}$ are three trilinear Yukawa couplings; $i,j,k = 1,2,3$
are generation indices; $L$ and $Q$ are the $SU(2)_L$ doublet superfields of the lepton and quark; $H_{\rm u}$
is the Higgs field that gives mass to the up-type quarks; $\mu_i$ are the bilinear terms that mix
lepton and Higgs superfields, and $\overline{E}$, $\overline{D}$, and $\overline{U}$ are the $SU(2)_L$ singlet
superfields of the charged lepton, down-type quark, and up-type quark. The third term violates the conservation
of baryon number, while the first two violate the conservation of lepton number. If baryon number and lepton
number were both violated, proton decay would proceed at a rate excluded by experimental
observations~\cite{protonDecay1, protonDecay2}. To avoid these experimental constraints and to simplify
the interpretation of results, it is commonly assumed that only one of the  $\Lam_{ijk}$, $\Lamp_{ijk}$,
or $\Lampp_{ijk}$ couplings is different from zero. In this analysis only $\Lamp_{ijk}$ couplings with $(i,j,k)\leq$ 2 are considered.

\begin{figure}[htbp]
 \centering
 \includegraphics[width=\cmsFigWidth]{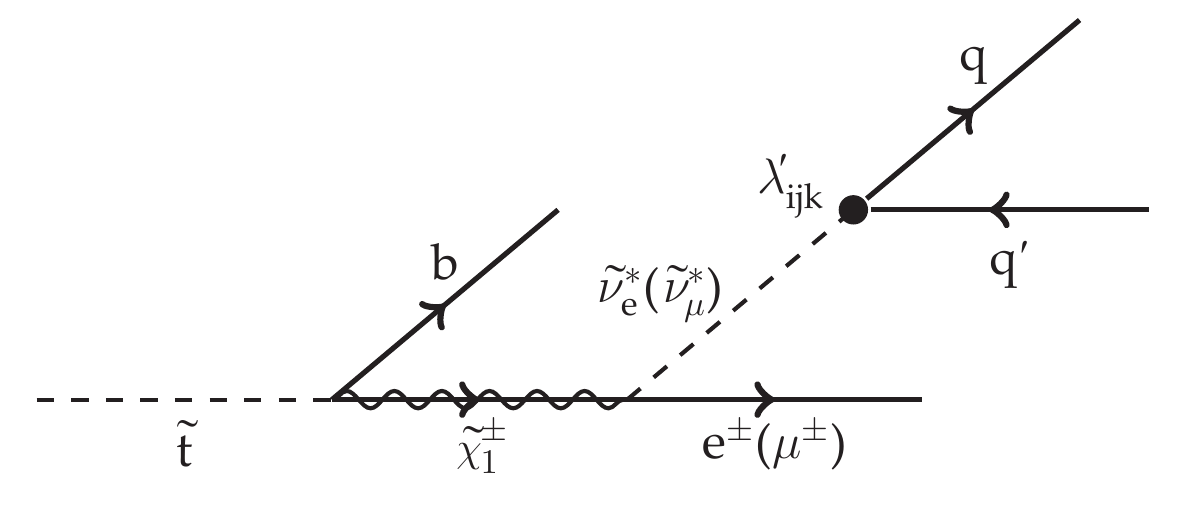}
 \caption{Diagram for the R-parity violating, chargino-mediated decay of a top squark. The chargino decays to a lepton
          and two jets via an off-shell sneutrino with nonzero $\Lamp_{ijk}$ coupling.}
 \label{Fig:decaychain}
\end{figure}

In RPV SUSY models with the chargino $\PSGcpmDo$ lighter than the top squark and nonzero $\Lamp_{ijk}$,
the top squark $\PSQt$ can decay via $\PSQt\to \PQb \PSGcpmDo$,
with subsequent decay of the chargino to a lepton and two jets via an off-shell sneutrino
($\PSGcpmDo\to \ell^\pm+\mathrm{jj}$)~\cite{EvansRPV}, as depicted in Fig.~\ref{Fig:decaychain}.
The branching fraction of decay $\PSGcpmDo\to \nu+\mathrm{jj}$ via an off-shell slepton will be
negligible unless the slepton and sneutrino masses are comparable. The decay
$\PSGcpmDo\to \PW^\pm \PSGczDo$ is suppressed for models with $\PSGcpmDo$
and $\PSGczDo$ almost degenerate in mass.

We perform a search for top squark decays, as depicted in Fig.~\ref{Fig:decaychain}, using proton-proton (\Pp\Pp)
collisions at a center-of-mass energy of 8\TeV, corresponding to an integrated luminosity of 19.7\fbinv, collected with
the CMS detector at the LHC in 2012. As top squarks are expected to be dominantly pair-produced at the LHC~\cite{squarkPair},
the search is performed using events with exactly two oppositely charged electrons (\ee) or muons (\mumu), at least five
jets of which one or more jet is identified as arising from hadronization of a bottom quark (b-tagged jet), and high \st,
where \st\ is defined as the scalar
sum of the transverse momenta of leptons and jets. As a consequence of the assumption that only one of the $\Lamp_{ijk}$
couplings is nonzero, the two leptons must have opposite charge and the same flavor. Details of the event selection are
described in Section~\ref{sec:sele}.

The sensitivity of the \ee (\mumu) search does not depend on which of the four RPV couplings associated with the second
operator LQD ($L_iQ_j\overline{D}_k$) in Eq.~(\ref{eqn:wrpv}): $\Lamp_{111}$, $\Lamp_{112}$, $\Lamp_{121}$, and $\Lamp_{122}$
($\Lamp_{211}$, $\Lamp_{212}$, $\Lamp_{221}$, and $\Lamp_{222}$) is non-zero, because the final states and kinematic
distributions are the same in each case. We expect that the searches have some sensitivity to models with third-generation
couplings $\Lamp_{311}$, $\Lamp_{312}$, $\Lamp_{321}$, and $\Lamp_{322}$, via leptonic $\tau$ decays; however, we do not
include this possible extra contribution in this paper. The difference \dm\ between top squark mass \mstop, and chargino
mass \mchi, is chosen to be 100\GeV, since this value is representative of the bulk of the \mstop-\mchi parameter
space where the signal reconstruction efficiency is slowly varying. This analysis does not attempt to quantify the decrease in
efficiency (and signal sensitivity) in the regions of parameter space where either \dm or \mchi is very small ($<$100\GeV).

Several searches for R-parity violating top squark decays via LQD couplings have been performed by the
CMS \cite{RPVstopSUS13003, EXO12032, EXO12041} and ATLAS \cite{ATLAS-CONF-2015-015} collaborations.
These searches have focused on top squark pairs decaying
via $\Lamp_{i32}$ couplings into final states of two leptons ($\Pe^\pm$ or $\PGm^\pm$) and two jets or two leptons
($\Pe^\pm$ or $\PGm^\pm$) and six jets, four of which are b-tagged jets~\cite{EXO12041, ATLAS-CONF-2015-015};
via $\Lamp_{3jk}$ couplings into a final state including two tau leptons and two b-tagged jets~\cite{EXO12032};
and via the $\Lamp_{233}$ coupling into a final state including three leptons and additional jets~\cite{RPVstopSUS13003}.
The analysis described in this paper is the first search for R-parity violating top squark decays via purely first- or
second-generation LQD couplings; in this case, the final states are two leptons ($\Pe^\pm$ or $\PGm^\pm$) and six jets,
two of which are b-tagged jets.

\section{The CMS detector \label{sec:cms}}

A detailed description of the CMS detector, together with a definition of the coordinate system used, can be found
elsewhere~\cite{cmsdet}. A notable feature of the CMS detector is its 6 m internal diameter superconducting solenoid
magnet that provides a field of 3.8\unit{T}. Within the field volume are a silicon pixel and strip tracker, a lead tungstate
crystal electromagnetic calorimeter, and a brass and scintillator hadron calorimeter. Muon detectors based on gas
ionization chambers are embedded in a steel flux-return yoke located outside the solenoid. Events are collected by a
two-layer trigger system based on a hardware level-1 trigger, followed by a software-based high-level trigger.

The pseudorapidity range covered by the tracking system is $\abs{\eta} < 2.5$, the muon detector extends up to
$\abs{\eta} < 2.4$, and the calorimeters cover a region with $\abs{\eta} < 3.0$. The region of $3 < \abs{\eta} < 5$ is
instrumented with steel and quartz fiber forward calorimeters. The hermeticity of the detector up
to large values of $\abs{\eta}$ permits accurate measurement of the momentum balance transverse to the beam direction.

\section{Trigger and event selection}
\label{sec:sele}

Events are selected using a trigger that requires at least one electron (muon) with a transverse momentum
(\pt) threshold of 27 (24)\GeV, and $\abs{\eta} < $ 2.5 (2.1).
All objects are reconstructed using a particle-flow (PF) algorithm~\cite{CMS:2010byl, CMS-PF},
which uses information from all subsystems to reconstruct photons, electrons, muons, charged hadrons, and neutral hadrons.

To reduce the background from jets containing leptons, we impose isolation constraints on the transverse energy
$\ETcone$ from charged-particle tracks or deposits in the calorimeter within a cone
$\Delta R = \sqrt{\smash[b]{(\Delta\eta)^{2} + (\Delta\phi)^{2}}} = 0.3\,(0.4)$ around the trajectory of the electron (muon),
where $\phi$ is the azimuthal angle. The energy from the reconstructed lepton
and the average transverse energy density from pileup are subtracted from $\ETcone$, where pileup is defined as additional
inelastic \Pp\Pp\xspace collisions within the same or the adjacent LHC bunch crossing. Tracking information together with calorimeter
information is used to identify and subtract hadronic energy depositions from charged particles originating from pileup.
The contributions to the neutral hadron and photon energy components due to pileup are also computed and subtracted.
In the electron channel, the contributions to the neutral hadron and photon energy components due to pileup interactions are
subtracted from $\ETcone$ using the jet area technique~\cite{Cacciari:2007fd}, which computes the transverse energy
density of neutral particles from the median of the neutral energy distribution in jets with $\pt>3\GeV$ on an
event-by-event basis. In the muon channel, the method assumes the pileup energy density from neutral particles to be half
of that from charged hadrons, based on measurements performed in jets~\cite{CMS-PF}.

Electrons are reconstructed by matching an energy cluster in the ECAL with a track reconstructed using a Gaussian
sum filter~\cite{Khachatryan:2015hwa}. Electrons are required to have $\pt>50\GeV$ and $\abs{\eta} < 2.5$. The
transition region between the ECAL barrel and endcap is excluded ($1.444  < \abs{\eta} < 1.566$) because the
calorimeter is not well modeled in this region. Electrons are identified using a multivariate identification
algorithm~\cite{Khachatryan:2015hwa}, whose input variables are sensitive to bremsstrahlung along the electron path,
matching between tracks and ECAL energy deposits, and shower-shape variables. The algorithm is trained with a sample of
simulated Drell--Yan (DY) events that contains true electrons and a data sample enriched in misidentified electrons. In
addition, the transverse impact parameter of the electron track is required to be less than 2\unit{mm}. To reduce backgrounds
that arise from photon conversions in the inner pixel detector, at least one pixel hit in the innermost pixel layer is
required and the electron must be inconsistent with the hypothesis that it resulted from photon pair creation. We ensure that
the electron is isolated from other activity in the event by requiring that $\ETcone$ be less than 10\% of the
electron \pt.

Muon tracks are reconstructed using the information from the muon chambers and the silicon tracker and are required to be consistent with
the reconstructed primary vertex. The tracks are required to have at least one hit in both the pixel tracker and muon detector,
and at least six hits in the silicon strip tracker. Muons are required to have $\pt > 50$\GeV and $\abs{\eta} < 2.1$. Most cosmic
ray muons are rejected by requiring that the transverse (longitudinal) impact parameter be less than 2\,(5)\unit{mm} relative to
the primary vertex, defined as the vertex with the largest sum of the $\pt^{2}$ from all tracks associated with it. Only muons with
at least ten hits in the silicon strip tracker and at least one hit in the pixel detector are considered, which ensures a precise
momentum measurement. Isolation is imposed by the requirement that $\ETcone$ be less than 12\% of the muon \pt~\cite{MuonIso}.

The differences in lepton reconstruction and trigger efficiencies between data and simulation are corrected in
simulation in bins of \pt and $\eta$, using a tag-and-probe method~\cite{TagAndProbe}.

Jets are reconstructed from PF objects~\cite{CMS:2009nxa} using the anti-\kt clustering algorithm~\cite{Cacciari:2008gp}
with a distance parameter of 0.5. The tracker and ECAL granularity are exploited to precisely measure the charged particles,
and hence to determine jet directions at the production vertex. To remove jets arising from instrumental and non-collision
backgrounds, additional criteria on charged and neutral hadron energy are applied.

The energy and momentum of each jet are corrected as a function of the jet \pt and $\eta$ to account for the combined
response function of the calorimeters. The average energy from pileup is subtracted from the jet~\cite{CMSJes}.
Only jets within $\abs{\eta} < 2.4$
are considered. The corrected jet \pt must be at least 100\GeV for the leading jet, 50\GeV for the second-leading jet, and 30\GeV
for the remaining jets. At least five jets are required in the event.

Events with at least one b-tagged jet are selected. The combined secondary-vertex algorithm~\cite{Chatrchyan:2012jua}
uses information from the track impact parameter and vertex information to discriminate between jets that originate from b
quarks and jets from light-flavor quarks and gluons. The algorithm correctly identifies jets produced by the hadronization
of a b quark (b jets) with an efficiency of approximately 70\% and misidentifies jets from light-flavor quarks or gluons
(charm quarks) at a rate of approximately 1\%\,(20\%)~\cite{Chatrchyan:2012jua}. The b-tagging efficiency in the simulation
is scaled to match the measured efficiency in data as a function of \pt, $\eta$, and the flavor of the jet.

The missing transverse momentum \ptvecmiss in the event is defined as the projection of the negative vector sum of the
momenta of all reconstructed PF candidates on the plane perpendicular to the beams. The magnitude of \ptvecmiss in the event is referred to as \ETm. To suppress leptonic \ttbar decays that often have significant
\ETm because of the presence of neutrinos in the final state, \ETm is required to be less than 100\GeV.
The dilepton mass \mll, computed from the two lepton four-momenta, is required to be greater
than 130\GeV, based on an optimization to reduce the contribution from low-mass resonances and Z boson decays.

To enhance the statistical significance, for each lepton flavor the sample is divided into three exclusive categories of jet
multiplicity: $\nj = 5,$ 6, or $\geq$ 7. To improve the sensitivity to signal decays, we compute an \st threshold \stmin optimized for
each top squark mass hypothesis and for each \nj bin. The \stmin is determined by maximizing the value of ${\mathrm{S}/\sqrt{\mathrm{S}+\mathrm{B}}}$,
where S and B are the number of expected signal and background events above \stmin, respectively.

\section{Simulation of background and signal events}
\label{sec:mc}

Monte Carlo (MC) simulations of background and signal events are used to optimize the selection criteria
for maximum signal sensitivity and to estimate backgrounds. The simulation of the hard-scattering event is
performed using the leading-order (LO) matrix element event generator \MADGRAPH5 \cite{MADGRAPH}, unless
noted otherwise. The CTEQ6L1 \cite{cteq6l1} set of parton distribution functions (PDF) is used to describe
the proton structure. The simulation of the hard-scattering event is then passed to \PYTHIA 6.426 \cite{PYTHIA}
with the Z2* tune~\cite{Chatrchyan:2013gfi} to model the parton shower, hadronization,
and the underlying event. A full simulation of the response of the CMS detector is performed using \GEANTfour \cite{GEANT4}.
Additional simulated minimum bias events are overlaid to reproduce the effects of pileup.

The main SM backgrounds for this search are DY and \ttbar pair production. Additional SM backgrounds, which include diboson
(WW, WZ, and ZZ) and single top quark production, are small. The \ttbar sample is generated with up to three additional partons,
the DY events are produced with up to four additional partons, and the diboson samples are generated with up to two additional
partons. Single top quark production ($t$-, $s$-, and tW-channels) is simulated with
\POWHEG~v1.0~\cite{Alioli:2009je, Re:2010bp,Nason:2004rx, Frixione:2007vw, Alioli:2010xd}. Simulated samples of \ttbar and
DY are normalized using cross sections computed at next-to-next-to-leading-order (NNLO)~\cite{topnnlo, dynnlo}. Cross sections
computed at next-to-leading-order (NLO)~\cite{singletopnnlo, dibosonnlo} are used to normalize the single top quark and diboson samples.

The signal samples are generated using {\MADGRAPH5}, \PYTHIA 6.426, and the CTEQ6L1 PDF set.
The top squark pair production cross section is computed at NLO as a function of \mstop, including soft gluon resummation at
next-to-leading logarithm (NLL)~\cite{Beenakker:1996ch, Kulesza:2008jb, Kulesza:2009kq, Beenakker:2009ha}. The uncertainty in
the cross section includes uncertainties associated with the renormalization and factorization scale, and the PDF set~\cite{nloxsec}.

\section{Background estimation}
\label{sec:bkg_lepton}

Corrections to the normalization of \ttbar and DY simulations are estimated by examining background enriched samples in
data. A summary of the selection criteria for the signal search region and the control regions, including selections on the
dilepton mass, is presented in Table~\ref{Table:SearchCntrl}. Diboson and single top quark production yield small contributions
to the background and are estimated from simulation. In simulated \ttbar sample, events are reweighted so that the \pt of the
top quark matches the data in a dedicated control sample~\cite{TopPt2}.

\begin{table*}[htb]
\centering
\topcaption{Summary of the selection criteria for the signal region and the control regions. Data in the control regions
         described as \ttbar, DY normalization, and DY shape are used to estimate SM backgrounds in the signal
         region. \label{Table:SearchCntrl}}
\begin{tabular}{llccc}
\hline
\multicolumn{1}{c}{}&      & Lepton selection  &$N_\text{jets}$ & $N_{\PQb-\text{tags}}$ \\ \hline
\multicolumn{1}{c}{Signal region}         &      & \ee (\mumu), $M_{\ell\ell} >$130\GeV  &$\geq$5       & $\geq$1\\
\hline\\[-2.1ex]
               & \ttbar shape     & \emu              &$\geq$5 & $\geq$1\tabularnewline
Control regions& DY normalization & \ee (\mumu), $50  < M_{\ell\ell} < 130$\GeV&$\geq$5 &$\geq$1\\
               & DY shape         & \ee (\mumu), $50  < M_{\ell\ell} < 130$\GeV&$\geq$5 &0       \\
\hline
\end{tabular}
\end{table*}

The leptonic \ttbar decays contribute to 89\% of the total background. Since the signal produces only same-flavor leptons,
we estimate the \ttbar background from a control sample of \emu events after correcting it for the small contributions of DY,
diboson, and single top events using simulations. We use this control sample to compute correction factors for the \ttbar
simulation for different jet multiplicities in the signal region. The \emu control sample is well modeled by the simulation,
thus correction factors are statistically consistent with unity.

The Drell--Yan production constitutes approximately 8\% of the SM background in the signal region, and is reduced by requiring
at least one b-tagged jet. The contribution from this source is estimated using a control sample of two oppositely charged
same-flavor leptons, which have an invariant mass \mll in the range 50--130\GeV. We perform a fit to the \mll distribution to
estimate the number of DY events. The DY shape is obtained from background-subtracted data using a DY-enriched sample with no
b-tagged jets. The background from diboson decays including leptonic Z boson decays is estimated from simulation and is constrained
in the fit. The \mll shape for the remaining backgrounds does not exhibit a Z boson mass peak, and is described by a linear
function. The fit determines the number $N_\mathrm{DY}$ of DY events and the number of all other background events. To check that
the procedure is insensitive to a
potential signal contamination, we performed a fit with signal events included, and observed that the obtained $N_\mathrm{DY}$ is
independent of the presence of the potential signal in the control sample. The ratio of $N_\mathrm{DY}$ from the fit to the
simulated number of DY events is calculated for each value of \nj and is used to correct the simulation. This correction factor ranges
from $1.2\pm0.1$ to $2.1\pm0.6$ and increases with jet multiplicity.

We checked that the corrections to the DY normalization are valid in the signal region with $\mll>130$\GeV. We compared the
numbers of events in different mass ranges using a DY-enriched sample with at least five jets and no
b-tagged jets. The ratio of the number of events with \mll in the Z-peak (normalization region) to the number with \mll in
the high-mass tail (signal region) is predicted from simulation to be $11.8 \pm 0.4$ and observed to be $14.0\pm 3.5$ in data,
in reasonable agreement.

\section{Systematic uncertainties}
\label{sec:sysUnc}

We evaluate systematic uncertainties related to each background and to the signal reconstruction efficiency; these are summarized
in Table~\ref{tab:sys_summary}.

\begin{table}[htb]
\centering
\topcaption{Systematic uncertainties for background and expected signal yields.\label{tab:sys_summary}}
\cmsTableResize{\begin{tabular}{llc}
\hline
                                          & Source                      & Uncertainty~(\%)  \\
\hline\\[-2ex]
                                          & $\ttbar$+jets               & 10--50 \\
Background                                & Drell--Yan                  & 50--100 \\
estimates                                 & Diboson                     & 30 \\
                                          & Single top quark            & 30 \\
                                          & MC statistics               & 10--30 \\
\hline
& Jet energy scale               &     5  \\
                                          & b tagging scale factor         &     1--3 \\
                                          & Integrated luminosity          &     2.6 \\
Expected                                  & Lepton identification          &     3 \\
signal yield                              & Electron energy scale          &     2 \\
                                          & Muon momentum scale            &     0.9 \\
                                          & Trigger efficiency             &     1 \\
                                          & Lepton isolation               &     5 \\
                                          & MC statistics                  &     2--7 \\
\hline
\end{tabular}}
\end{table}

Since the \ttbar correction factor for the simulated sample is estimated from a control sample of \emu events in data,
the systematic uncertainty in this background is given by the statistical uncertainty in the control sample. This uncertainty
ranges from 10 to 50\%, depending on the value of \nj and of \stmin. The uncertainties related to lepton trigger, identification,
and isolation are negligible. For the small DY background, we take 50\,(100)\% of correction factor as the systematic uncertainty
on the correction in 5 ($\geq$6) jet bin(s). We assign a 30\% uncertainty to the diboson and single top quark background
contributions to account for the difference between the NLO theoretical calculation and the CMS measurements of the WW and ZZ
cross sections~\cite{cmsdiboson} and the single top cross sections~\cite{cmsSingleTop_t}. The statistical uncertainty due to the
finite size of the simulated background samples is 10--30\%, depending on the \nj bin and \stmin value.

The following systematic uncertainties in the signal efficiency are included: jet energy scale (5\%)~\cite{CMSJes},
jet b-tagging efficiencies (3\%), integrated luminosity (2.6\%)~\cite{lumi}, lepton identification and reconstruction
efficiency (3\%), electron energy scale (2\%), muon momentum scale (0.9\%), and trigger efficiency (1\%). Noting the
effect of the b-tagging uncertainty on the signal prediction is evaluated by varying the efficiency and misidentification
rates by their uncertainties~\cite{Chatrchyan:2012jua, CMS:2013vea}, and the effect on the signal prediction. The
uncertainty related to the lepton isolation requirement for signal events with many jets is estimated using a \ttbar
control sample selected as shown in Table~\ref{Table:SearchCntrl}, but with $\geq$ 7 jets, and is determined to be 5\%.
The uncertainty due to the limited size of the simulated signal sample varies from 2 to 7\%. The impact of uncertainties
related to the PDF set choice, modeling of the top quark \pt spectrum, and pileup modeling is determined to be negligible.

\section{Results}
\label{sec:res}
\begin{figure*}[htbp]
 \centering
  \includegraphics[width=0.49\textwidth]{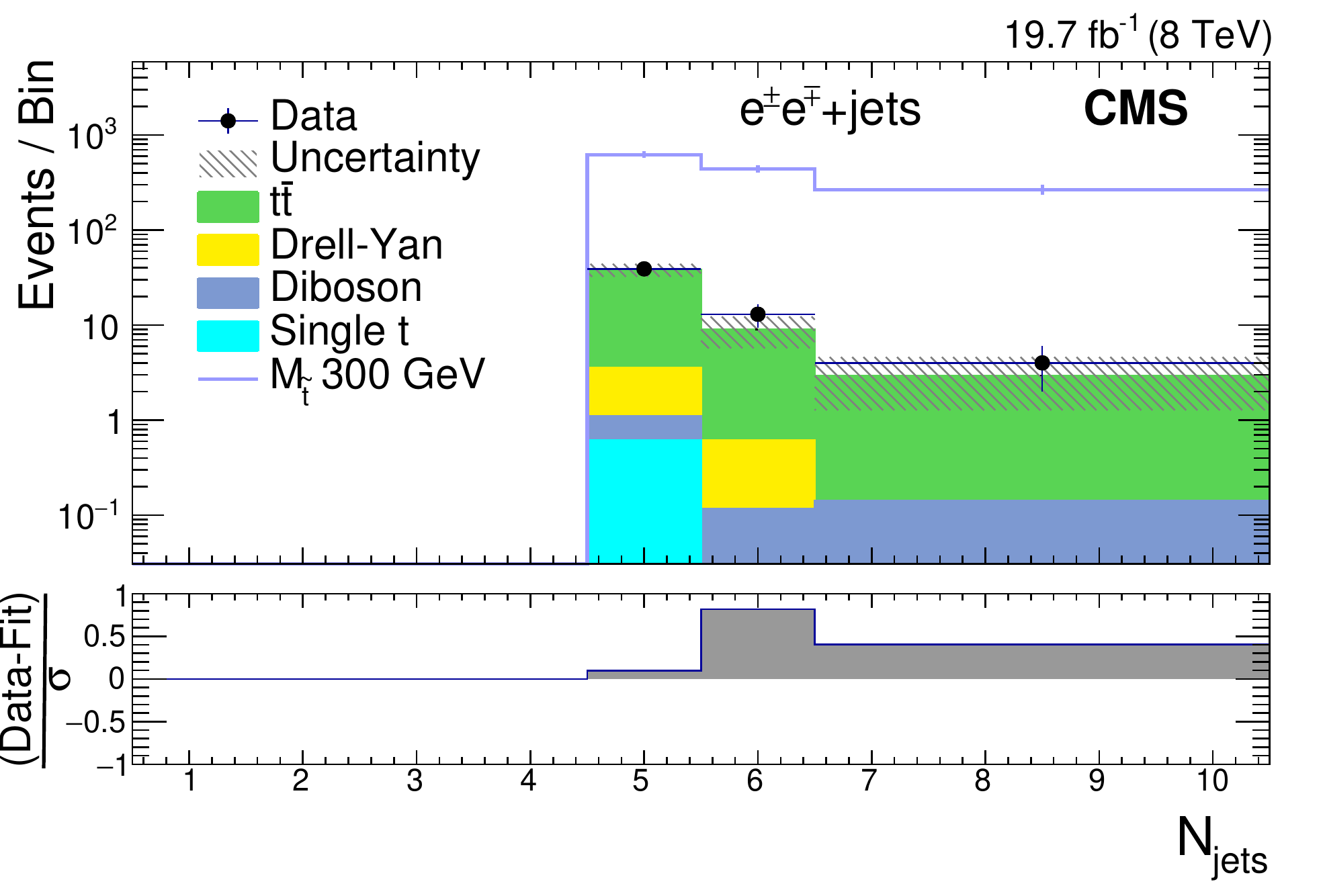} \includegraphics[width=0.49\textwidth]{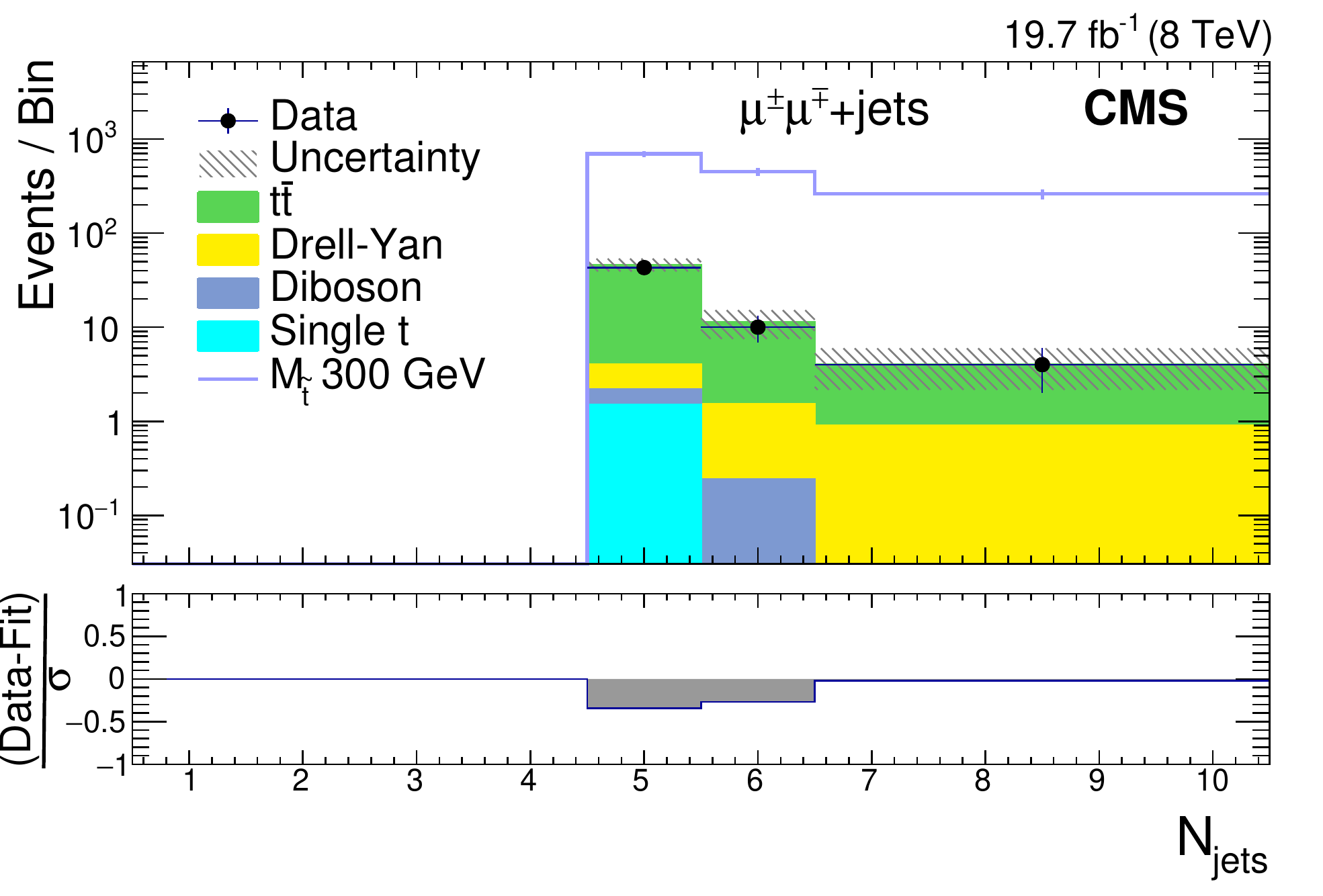}
  \includegraphics[width=0.49\textwidth]{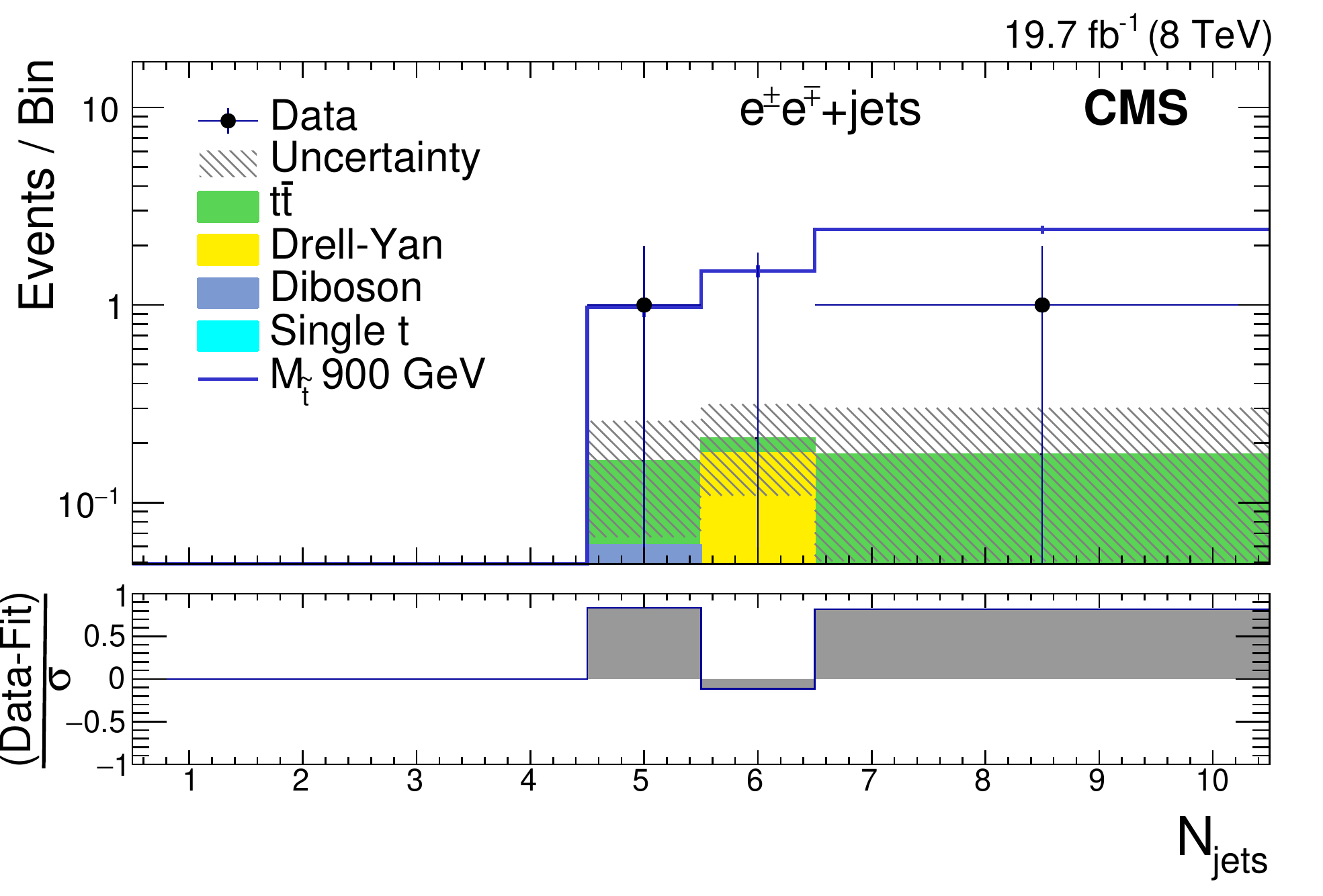} \includegraphics[width=0.49\textwidth]{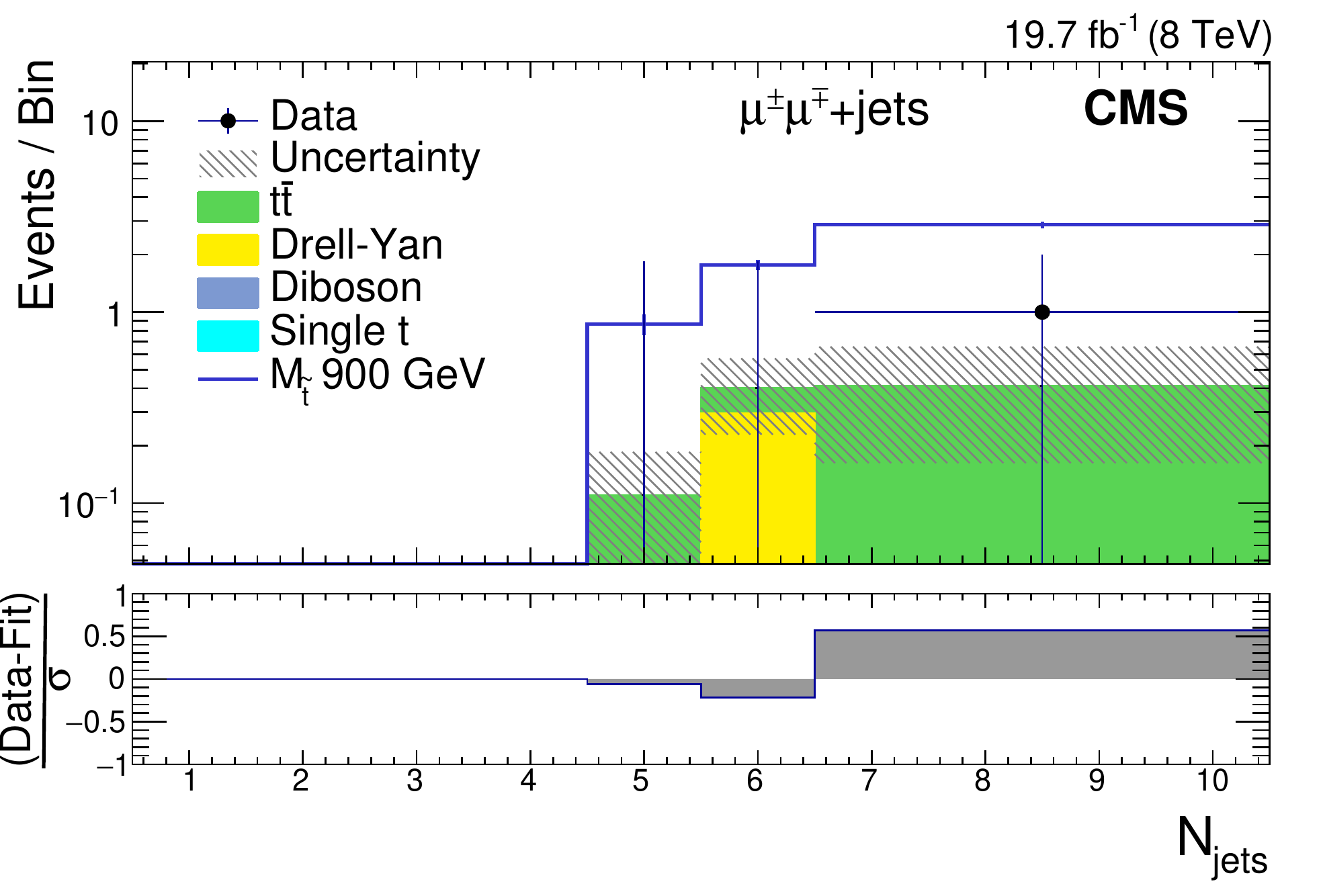}
  \caption{Jet multiplicity distributions for \ee (left) and \mumu (right) for selections optimized for
           \mstop hypotheses of 300\GeV (top) and 900\GeV (bottom). The expected signal is shown by an open histogram
           superimposed on the expected SM background. The asymmetric error bars indicate the central confidence intervals
           for Poisson-distributed data. 
           The systematic uncertainties for the SM contributions are indicated
           by hatched bands. Under each histogram is shown a plot in gray as the ratio of difference of data from
           background expectation to the sum of their uncertainties, including the systematic uncertainties in background
           expectation.
}
\label{fig:signalregion}
\end{figure*}

Figure~\ref{fig:signalregion} shows the observed distributions of jet multiplicity, the estimated background
distributions, and the expected distributions for signals with a mass \mstop of either 300\GeV or 900\GeV.
In Tables~\ref{tab:eleCounts} and \ref{tab:muCounts} we present the numbers of expected and observed events
for each value of \nj, for each \mstop\ hypothesis and corresponding \stmin value. The signal expectations are
based on NLO cross sections~\cite{nloxsec}. The data are in agreement with the SM expectation
in each bin. The corresponding distributions are displayed graphically in \suppMaterial.

We use these results to determine 95\% confidence level (CL) limits, as a function of \mstop, on the product of the
top-squark pair-production cross section and the square of the branching fraction $\mathcal{B}$ for the decay
$\PSQt\to \PQb\ell^{\pm}\PQq\PQq$. We use the modified frequentist CL$_\mathrm{s}$ method~\cite{CMS-NOTE-2011-005} with
profiling of nuisance parameters. For each \mstop hypothesis, the Poisson likelihoods of the three \nj bins are combined.
Systematic uncertainties are incorporated into the test statistic as nuisance parameters. The nuisance parameter probability
density function (pdf) for the \ttbar background normalization, which is estimated from background control regions containing
limited numbers of events in high \nj bins, is described by a gamma function. All other uncertainties are treated with
log-normal pdfs. With the exception of uncertainties related to the finite size of a control sample, we assume the systematic
uncertainties are fully correlated across different \nj bins.

\begin{figure*}[htbp]
  \centering
  \includegraphics[width=0.49\textwidth]{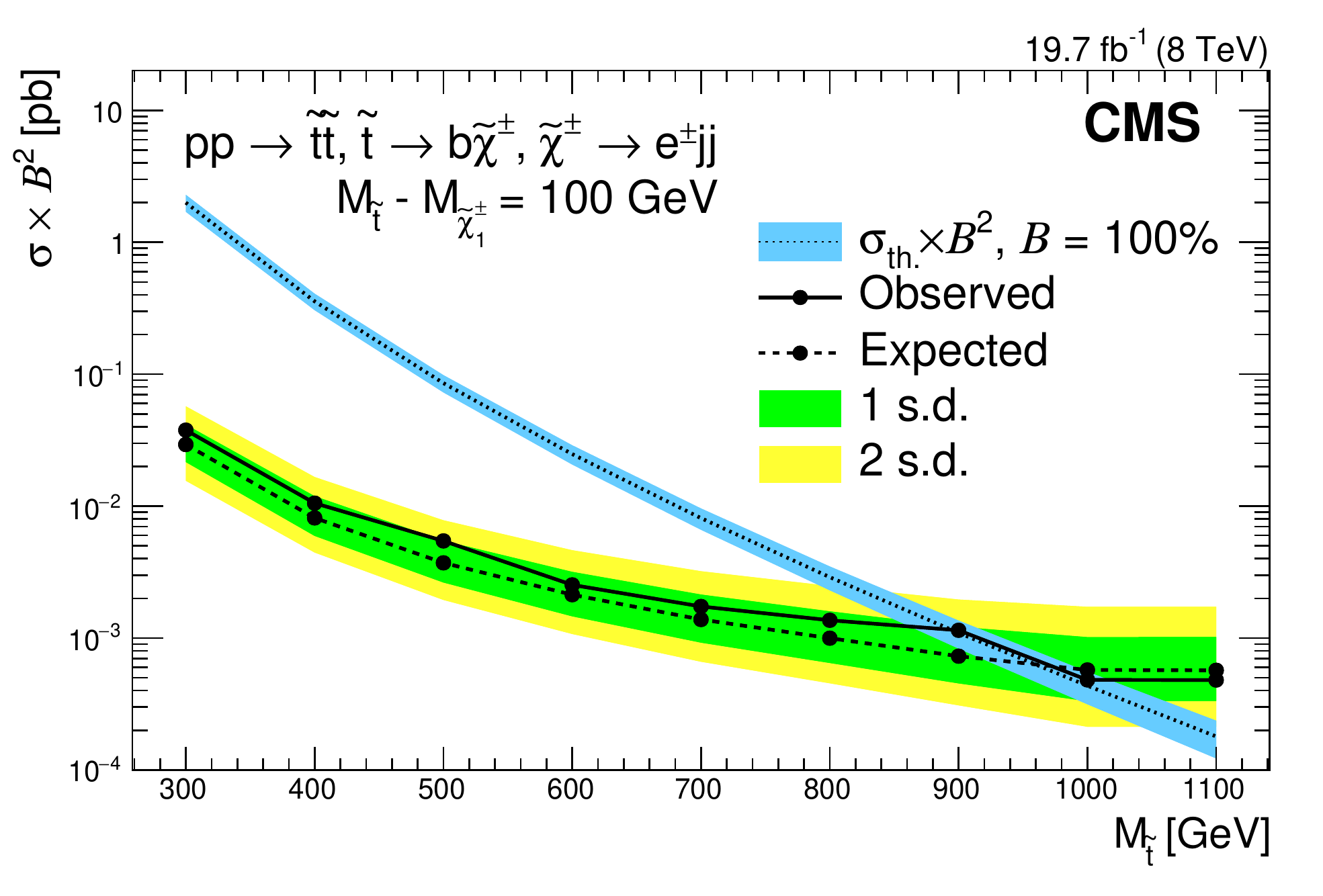} \includegraphics[width=0.49\textwidth]{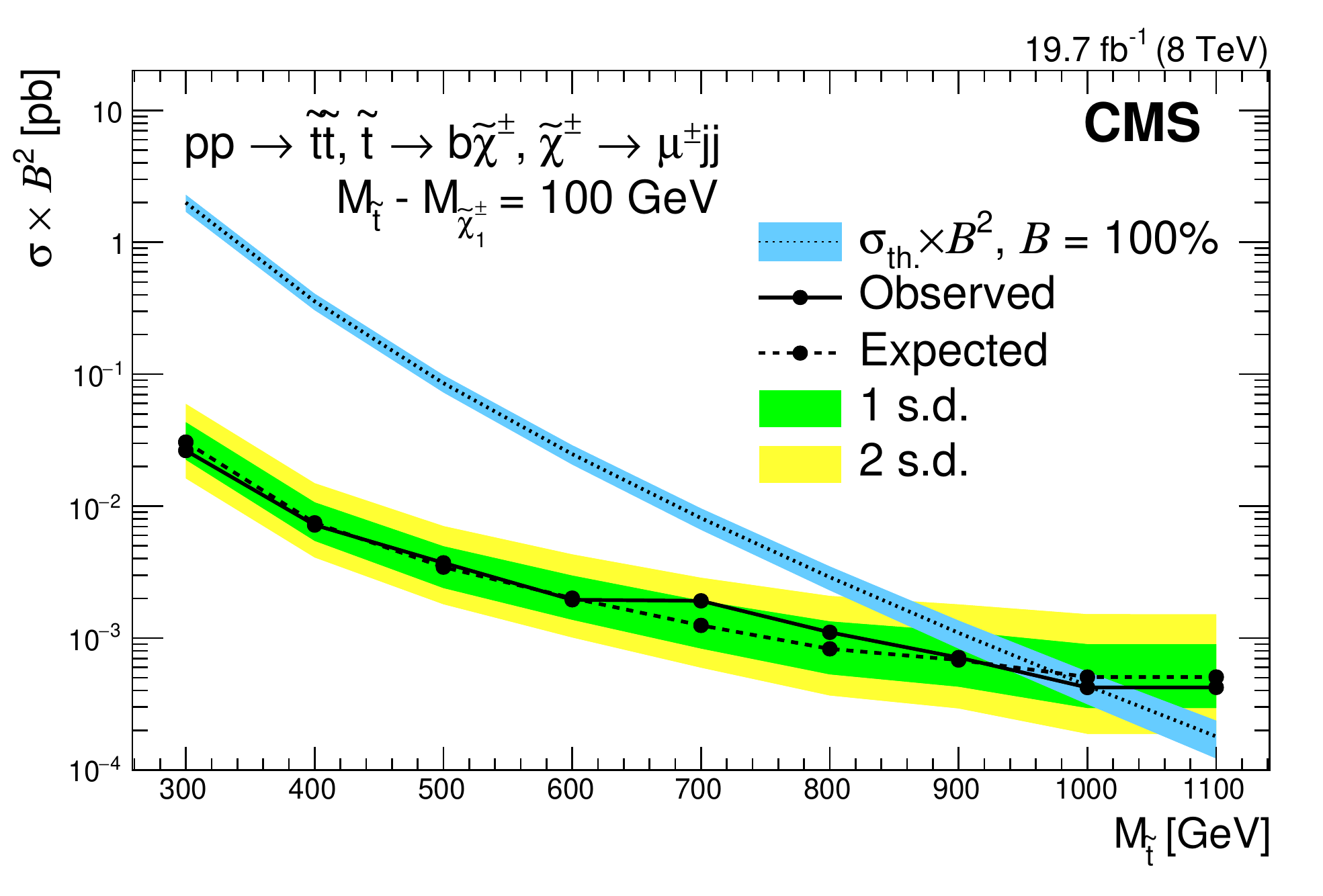}
  \caption{Observed and expected 95\% CL upper limits on the product of the cross section and the branching fraction
           ($\mathcal{B}$) squared, for \ee (left) and \mumu (right). The green (inner) and yellow (outer) bands show
           the 1 s.d. and 2 s.d. uncertainty ranges in the expected limits, respectively. The dotted curve shows the
           expected top squark cross section computed at NLO+NLL. The difference \mstop-\mchi\ is assumed to be 100\GeV
           for the signal model.}
             \label{fig:limitplot}
\end{figure*}

The observed and expected limits on the product of the cross section and the branching fraction squared are shown in
Fig.~\ref{fig:limitplot}. The green (yellow) band corresponds to a variation of one (two) standard deviation(s) on the
expected limit.  The dotted curve shows the signal cross section, with the width of the associated band showing the
sensitivity to uncertainties in the renormalization and factorization scales and the PDF uncertainties~\cite{nloxsec}.
Comparing the observed cross section limits to the signal cross section, we exclude top squarks with masses less than
890\,(1000)\GeV for the electron (muon) channel. The expected mass exclusion is 950\,(970)\GeV for the electron (muon) channel.

These cross section limits strictly apply to models with mass difference $\dm=100\GeV$; however, the sensitivities for models
with $\dm>50\GeV$ are similar. The mass exclusions assume $\mathcal{B}=100\%$. As described earlier, the limits for the electron
channel apply equally to models with nonzero $\Lamp_{111}$, $\Lamp_{112}$, $\Lamp_{121}$, or $\Lamp_{122}$ and the limits for
the muon channel apply equally to models with nonzero $\Lamp_{211}$, $\Lamp_{212}$, $\Lamp_{221}$, or $\Lamp_{222}$. Because
the coupling strength does not affect the production cross section and the branching fraction is assumed to be 100\%, the value
of $\Lamp_{ijk}$ is not important as long as it is sufficiently large to ensure that the sneutrino decays promptly. For
coupling values smaller than $10^{-5}$, the decay lengths are of order 1\unit{mm} or greater, resulting in a decreased
signal reconstruction efficiency and sensitivity. These are the first limits on chargino-mediated top squark decays via a
single LQD coupling $\Lamp_{ijk}$ with $(i,j,k\leq 2)$.

\section{Summary}
\label{sec:sum}
A search for new phenomena using events with two oppositely charged electrons or muons, at least five jets,
with at least one b-tagged jet, and low missing transverse momentum has been performed. No excess over the
estimated background is observed. The results are interpreted in the framework of chargino-mediated, R-parity
violating top squark decays, assuming a 100\% branching fraction for the top squark decay chain,
$\PSQt\to \PQb \PSGcpmDo, \PSGcpmDo\to \ell^\pm+\mathrm{jj}$.
In models with a single nonzero $\Lamp_{ijk}$ coupling with $(i,j,k \leq 2)$, the results exclude top squarks
with mass less than 890 (1000)\GeV for the electron (muon) channel at 95\% confidence level.
These limits are the first obtained for this model.

\begin{table*}[htbp]
\centering
\topcaption{Observed events, estimated background, and expected signal yields, for $\nj = 5,$ 6, and $\geq$7, along
         with the optimized value of \stmin, for different \mstop\ in the electron channel. The signal and background
         uncertainties include both statistical and systematic contributions.\label{tab:eleCounts} }
\begin{tabular}{r r r r D{,}{\pm}{-1} D{,}{\pm}{-1} D{,}{\pm}{-1}}
\hline

\multicolumn{1}{c}{\mstop (\GeVns{})}& \multicolumn{1}{c}{\nj} & \multicolumn{1}{c}{\stmin (\GeVns{})}& \multicolumn{1}{c}{Data} & \multicolumn{1}{c}{\text{Estimated}} & \multicolumn{1}{c}{\text{Expected}} & \multicolumn{1}{c}{\text{Signal}}  \\
                                &     &              &      & \multicolumn{1}{c}{\text{background}}& \multicolumn{1}{c}{\text{signal}}   & \multicolumn{1}{c}{\text{efficiency(\%)}}  \\ \hline

     300 & 5        & 325     & 39      & 38.1 , 5.9    &   622  , 49      & 2.4  , 0.2   \\
     300 & 6        & 325     & 13      &  9.0 , 3.3    &   442  , 41      & 1.8  , 0.1   \\
     300 & $\geq$7  & 325     & 4       &  2.9 , 1.7    &   266  , 33      & 0.9  , 0.1   \\
\hline
     400 & 5        & 525     & 27      & 28.7 , 5.6    &   256  , 14      & 5.6  , 0.2   \\
     400 & 6        & 325     & 13      &  9.0 , 3.3    &   245  , 13      & 5.3  , 0.2   \\
     400 & $\geq$7  & 325     & 4       &  2.9 , 1.7    &   180  , 11      & 3.8  , 0.2   \\
\hline
     500 & 5        & 725     & 12      & 14.1 , 3.3    &  69.2  , 3.3     & 6.0  , 0.2   \\
     500 & 6        & 675     & 9       &  5.3 , 2.5    &  88.1  , 3.7     & 7.9  , 0.3   \\
     500 & $\geq$7  & 675     & 4       &  2.2 , 1.4    &  89.7  , 3.8     & 8.1  , 0.3   \\
\hline
     600 & 5        & 925     & 1       &  3.4 , 1.1    &  19.0  , 0.9     & 5.8  , 0.2   \\
     600 & 6        & 875     & 3       &  2.7 , 1.0    &  28.8  , 1.1     & 8.9  , 0.3   \\
     600 & $\geq$7  & 825     & 4       &  1.8 , 0.9    &  38.7  , 1.3     &11.6  , 0.3   \\
\hline
     700 & 5        & 1025    & 1       &  1.6 , 0.5    &   7.1  , 0.3     & 6.6  , 0.2   \\
     700 & 6        & 975     & 2       &  1.3 , 0.5    &  10.5  , 0.4     & 9.6  , 0.3   \\
     700 & $\geq$7  & 975     & 2       &  1.1 , 0.6    &  14.8  , 0.5     & 13.6  , 0.3   \\
\hline
     800 & 5        & 1225    & 1       &  0.4 , 0.2    &   2.7  , 0.1     & 7.0  , 0.2   \\
     800 & 6        & 1175    & 0       &  0.4 , 0.2    &   3.6  , 0.2     & 9.5  , 0.3   \\
     800 & $\geq$7  & 1075    & 2       &  0.7 , 0.4    &   5.7  , 0.2     & 15.1  , 0.4   \\
\hline
     900 & 5        & 1325    & 1       &  0.2 , 0.1    &   1.0  , 0.1     & 6.7  , 0.3   \\
     900 & 6        & 1375    & 0       &  0.2 , 0.1    &   1.5  , 0.1     &10.1  , 0.3   \\
     900 & $\geq$7  & 1375    & 1       &  0.2 , 0.1    &   2.4  , 0.1     &16.4  , 0.4   \\
\hline
    1000 & 5        & 1475    & 0       & 0.06 , 0.07    & 0.34  , 0.10    & 5.7  , 0.2   \\
    1000 & 6        & 1425    & 0       & 0.18 , 0.10    & 0.61  , 0.09    &10.6  , 0.3   \\
    1000 & $\geq$7  & 1525    & 0       & 0.05 , 0.06    & 0.98  , 0.09    &16.6  , 0.4   \\
\hline
    1100 & 5        & 1475    & 0       & 0.06 , 0.07    & 0.12  , 0.04    & 5.3  , 0.2   \\
    1100 & 6        & 1425    & 0       & 0.18 , 0.10    & 0.26  , 0.04    &11.2  , 0.3   \\
    1100 & $\geq$7  & 1525    & 0       & 0.05 , 0.06    & 0.42  , 0.04    &17.6  , 0.4   \\
\hline
\end{tabular}
\end{table*}

\begin{table*}[!htbp]
\centering
\topcaption{Observed events, estimated background, and expected signal yields, for $\nj = 5,$ 6, and $\geq$7, along with the
         optimized value of \stmin, for different \mstop in the muon channel. The signal and background uncertainties include both
         statistical and systematic contributions. \label{tab:muCounts}
}
\begin{tabular}{r  r r r D{,}{\pm}{-1} D{,}{\pm}{-1} D{,}{\pm}{-1}}
\hline

\multicolumn{1}{c}{\mstop (\GeV)}& \multicolumn{1}{c}{\nj} & \multicolumn{1}{c}{\stmin (\GeVns{})}& \multicolumn{1}{c}{Data} & \multicolumn{1}{c}{\text{Estimated}} & \multicolumn{1}{c}{\text{Expected}} & \multicolumn{1}{c}{\text{Signal}}  \\
                                &     &              &      & \multicolumn{1}{c}{\text{background}}& \multicolumn{1}{c}{\text{signal}}   & \multicolumn{1}{c}{\text{efficiency(\%)}}  \\ \hline

     300 & 5        & 475     & 43      & 46.4  , 7.2   &   696 , 52     &  2.5  , 0.2   \\
     300 & 6        & 475     & 10      & 11.3  , 3.8   &   450 , 43     &  1.7  , 0.1   \\
     300 & $\geq$7  & 325     & 4       &  4.1  , 1.9   &   261 , 33     &  0.9  , 0.1   \\
\hline
     400 & 5        & 525     & 39      & 36.8  , 7.2   &   266 , 13     &  5.4  , 0.2   \\
     400 & 6        & 525     & 10      & 10.8  , 3.9   &   281 , 14     &  5.3  , 0.2   \\
     400 & $\geq$7  & 325     & 4       &  4.1  , 1.9   &   223 , 12     &  4.3  , 0.2   \\
\hline
     500 & 5        & 725     & 16      & 16.0  , 3.8   &  81.1 , 4.0    &  6.3  , 0.3   \\
     500 & 6        & 675     & 9       &  7.3  , 3.2   & 114.4 , 4.8    &  8.8  , 0.3   \\
     500 & $\geq$7  & 675     & 3       &  3.1  , 1.6   & 101.8 , 4.5    &  8.3  , 0.3   \\
\hline
     600 & 5        & 875     & 5       &  5.2  , 1.5   &  23.7 , 1.1    &  6.6  , 0.3   \\
     600 & 6        & 825     & 5       &  4.6  , 1.6   &  36.0 , 1.3    & 10.0  , 0.3   \\
     600 & $\geq$7  & 825     & 2       &  2.4  , 1.0   &  44.2 , 1.5    & 12.3  , 0.3   \\
\hline
     700 & 5        & 1075    & 2       &  1.3  , 0.4   &   7.7 , 0.4    &  6.3  , 0.2   \\
     700 & 6        & 975     & 4       &  2.4  , 0.8   &  13.2 , 0.5    & 11.2  , 0.3   \\
     700 & $\geq$7  & 975     & 2       &  1.0  , 0.5   &  17.8 , 0.5    & 14.9  , 0.4   \\
\hline
     800 & 5        & 1175    & 0       &  0.9  , 0.3   &   2.9 , 0.2    &  6.8  , 0.3   \\
     800 & 6        & 1175    & 2       &  0.8  , 0.3   &   4.5 , 0.2    & 10.6  , 0.3   \\
     800 & $\geq$7  & 1125    & 1       &  0.4  , 0.3   &   7.3 , 0.2    & 17.6  , 0.4   \\
\hline
     900 & 5        & 1475    & 0       &  0.1  , 0.1   &   0.9 , 0.1    &  5.6  , 0.2   \\
     900 & 6        & 1325    & 0       &  0.4  , 0.2   &   1.8 , 0.1    & 11.0  , 0.3   \\
     900 & $\geq$7  & 1175    & 1       &  0.4  , 0.3   &   2.9 , 0.1    & 18.1  , 0.4   \\
\hline
    1000 & 5        & 1575    & 0       &  0.07 , 0.06  &   0.4 , 0.1    &  5.9  , 0.2   \\
    1000 & 6        & 1525    & 0       &  0.01 , 0.04  &   0.6 , 0.1    & 10.0  , 0.3   \\
    1000 & $\geq$7  & 1425    & 0       &  0.25 , 0.16  &   1.2 , 0.1    & 18.9  , 0.4   \\
\hline
    1100 & 5        & 1575    & 0       &  0.07 , 0.06  &   0.13 , 0.04  &  5.2  , 0.3   \\
    1100 & 6        & 1525    & 0       &  0.01 , 0.04  &   0.25 , 0.04  &  9.9  , 0.3   \\
    1100 & $\geq$7  & 1425    & 0       &  0.25 , 0.16  &   0.50 , 0.04  & 19.7  , 0.4   \\
\hline
\end{tabular}
\end{table*}
\clearpage
\begin{acknowledgments}
We congratulate our colleagues in the CERN accelerator departments for the excellent performance of the LHC and thank the technical and administrative staffs at CERN and at other CMS institutes for their contributions to the success of the CMS effort. In addition, we gratefully acknowledge the computing centers and personnel of the Worldwide LHC Computing Grid for delivering so effectively the computing infrastructure essential to our analyses. Finally, we acknowledge the enduring support for the construction and operation of the LHC and the CMS detector provided by the following funding agencies: BMWFW and FWF (Austria); FNRS and FWO (Belgium); CNPq, CAPES, FAPERJ, and FAPESP (Brazil); MES (Bulgaria); CERN; CAS, MoST, and NSFC (China); COLCIENCIAS (Colombia); MSES and CSF (Croatia); RPF (Cyprus); MoER, ERC IUT and ERDF (Estonia); Academy of Finland, MEC, and HIP (Finland); CEA and CNRS/IN2P3 (France); BMBF, DFG, and HGF (Germany); GSRT (Greece); OTKA and NIH (Hungary); DAE and DST (India); IPM (Iran); SFI (Ireland); INFN (Italy); MSIP and NRF (Republic of Korea); LAS (Lithuania); MOE and UM (Malaysia); CINVESTAV, CONACYT, SEP, and UASLP-FAI (Mexico); MBIE (New Zealand); PAEC (Pakistan); MSHE and NSC (Poland); FCT (Portugal); JINR (Dubna); MON, RosAtom, RAS and RFBR (Russia); MESTD (Serbia); SEIDI and CPAN (Spain); Swiss Funding Agencies (Switzerland); MST (Taipei); ThEPCenter, IPST, STAR and NSTDA (Thailand); TUBITAK and TAEK (Turkey); NASU and SFFR (Ukraine); STFC (United Kingdom); DOE and NSF (USA).

Individuals have received support from the Marie-Curie program and the European Research Council and EPLANET (European Union); the Leventis Foundation; the A. P. Sloan Foundation; the Alexander von Humboldt Foundation; the Belgian Federal Science Policy Office; the Fonds pour la Formation \`a la Recherche dans l'Industrie et dans l'Agriculture (FRIA-Belgium); the Agentschap voor Innovatie door Wetenschap en Technologie (IWT-Belgium); the Ministry of Education, Youth and Sports (MEYS) of the Czech Republic; the Council of Science and Industrial Research, India; the HOMING PLUS program of the Foundation for Polish Science, cofinanced from European Union, Regional Development Fund; the OPUS program of the National Science Center (Poland); the Compagnia di San Paolo (Torino); MIUR project 20108T4XTM (Italy); the Thalis and Aristeia programs cofinanced by EU-ESF and the Greek NSRF; the National Priorities Research Program by Qatar National Research Fund; the Rachadapisek Sompot Fund for Postdoctoral Fellowship, Chulalongkorn University (Thailand); the Chulalongkorn Academic into Its 2nd Century Project Advancement Project (Thailand); and the Welch Foundation, contract C-1845.
\end{acknowledgments}

\clearpage

\bibliography{auto_generated}
\ifthenelse{\boolean{cms@external}}{}{
\clearpage
\newpage
\numberwithin{figure}{section}
\appendix

\section{Supplementary Material} \label{app:suppMat}

Figure~\ref{fig:stmin} shows the number of observed and predicted events in each optimized \stmin selection for 5th, 6th, and $\geq$ 7th jets for electron and muon channels. The event counts correspond to Tables~\ifthenelse{\boolean{cms@supplement}}{3}{\ref{tab:eleCounts}} and~\ifthenelse{\boolean{cms@supplement}}{4 in the main document,}{\ref{tab:muCounts}} respectively.

\begin{figure*}[htbp]
 \centering
  \includegraphics[width=0.49\textwidth]{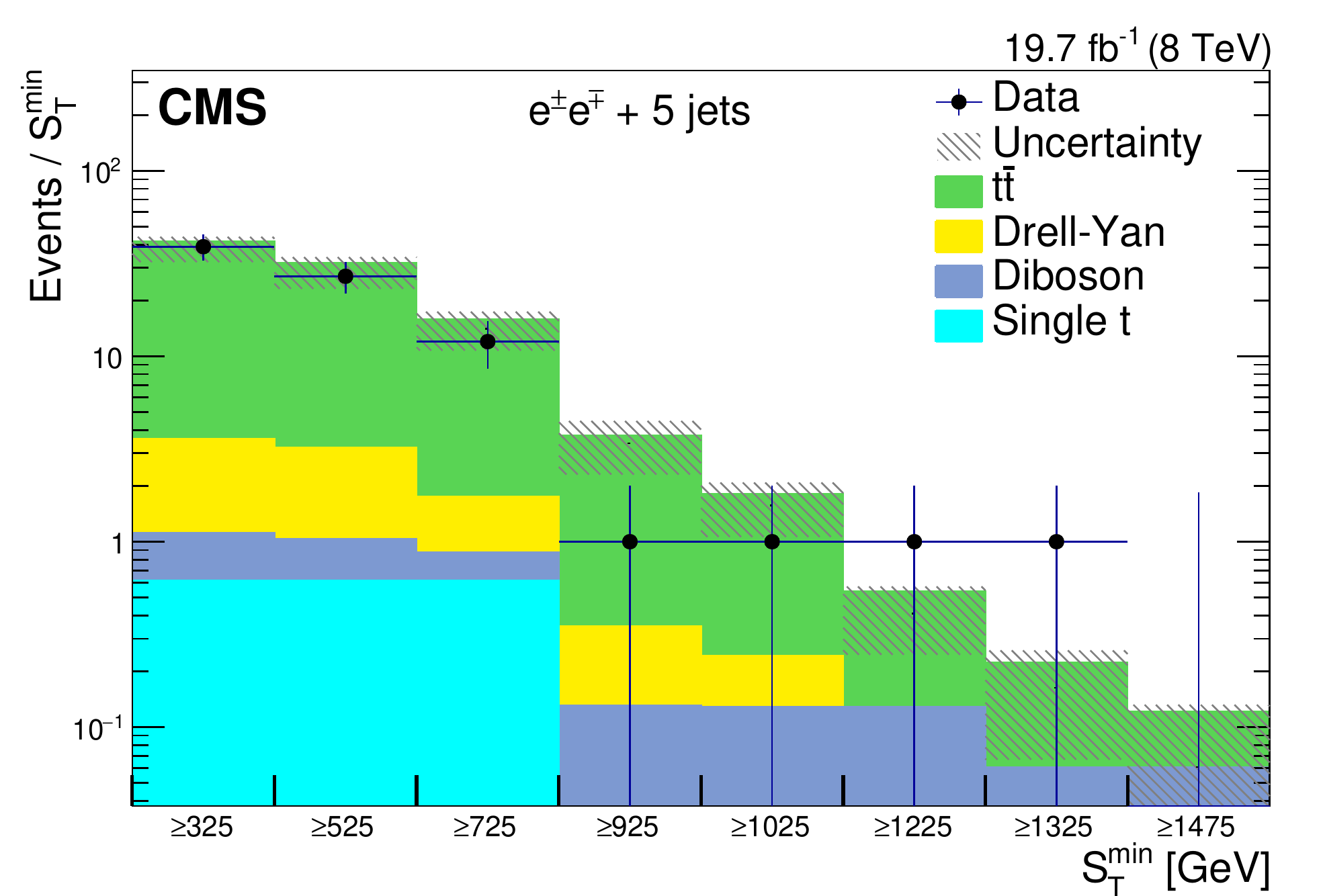}  \includegraphics[width=0.49\textwidth]{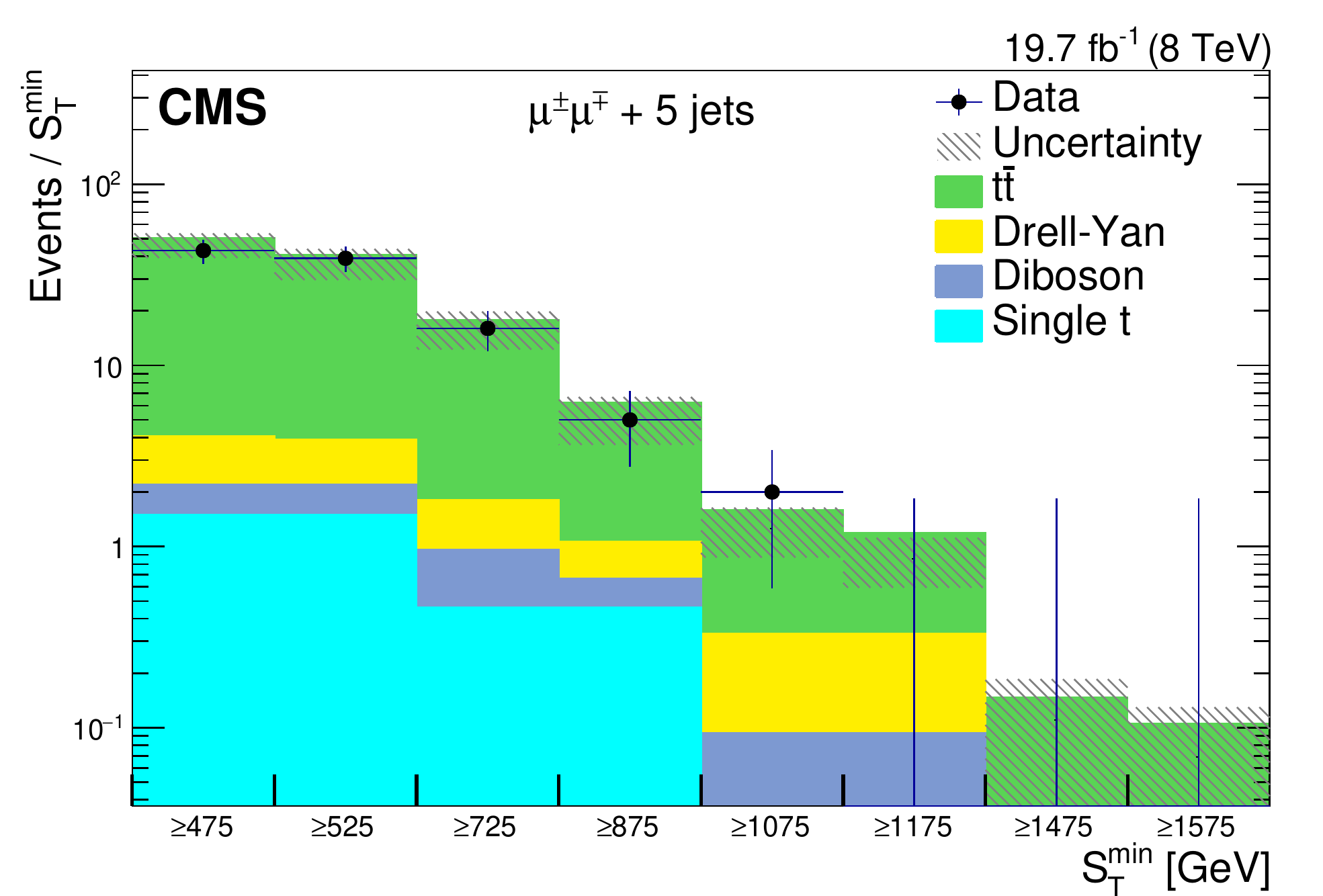}
  \includegraphics[width=0.49\textwidth]{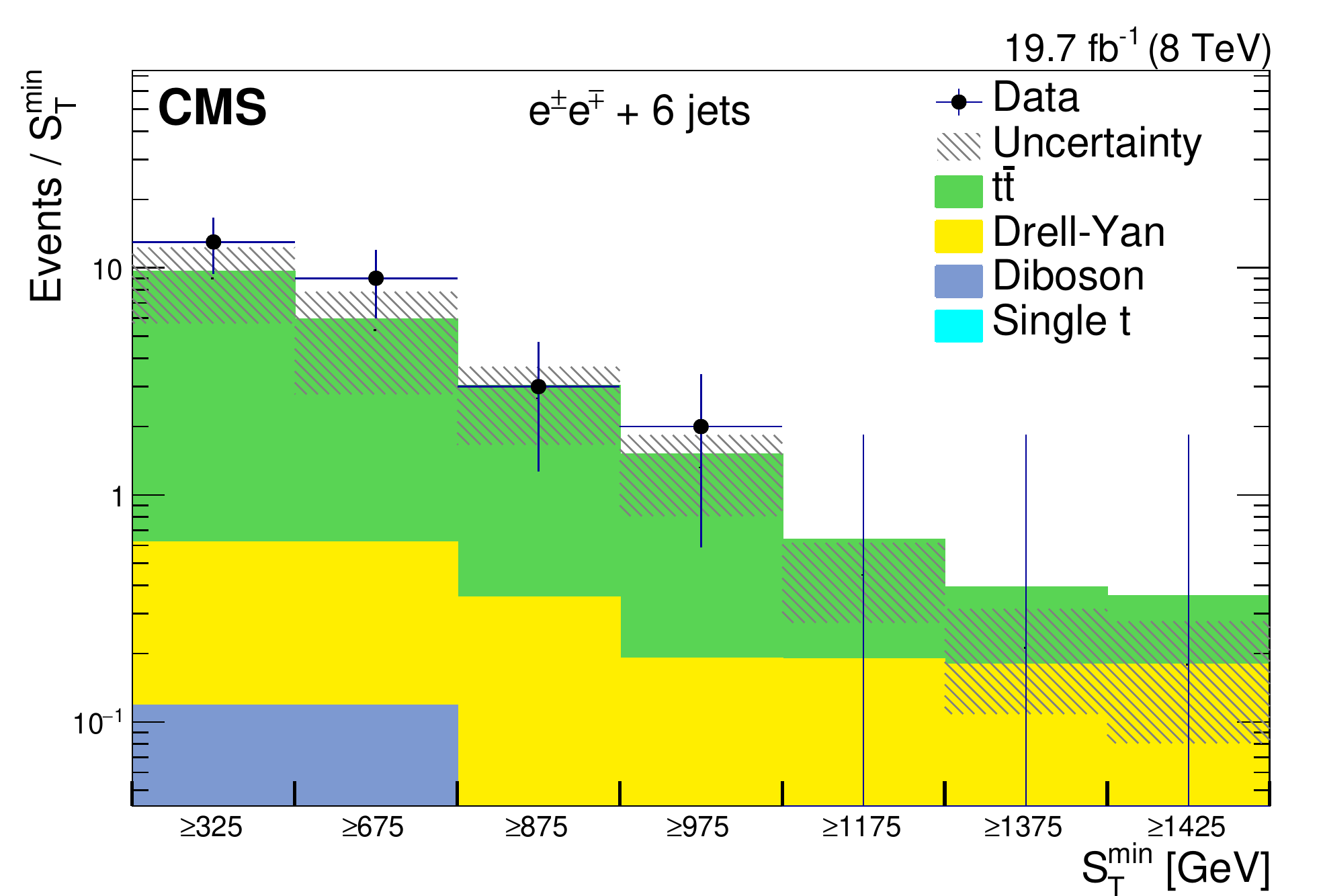} \includegraphics[width=0.49\textwidth]{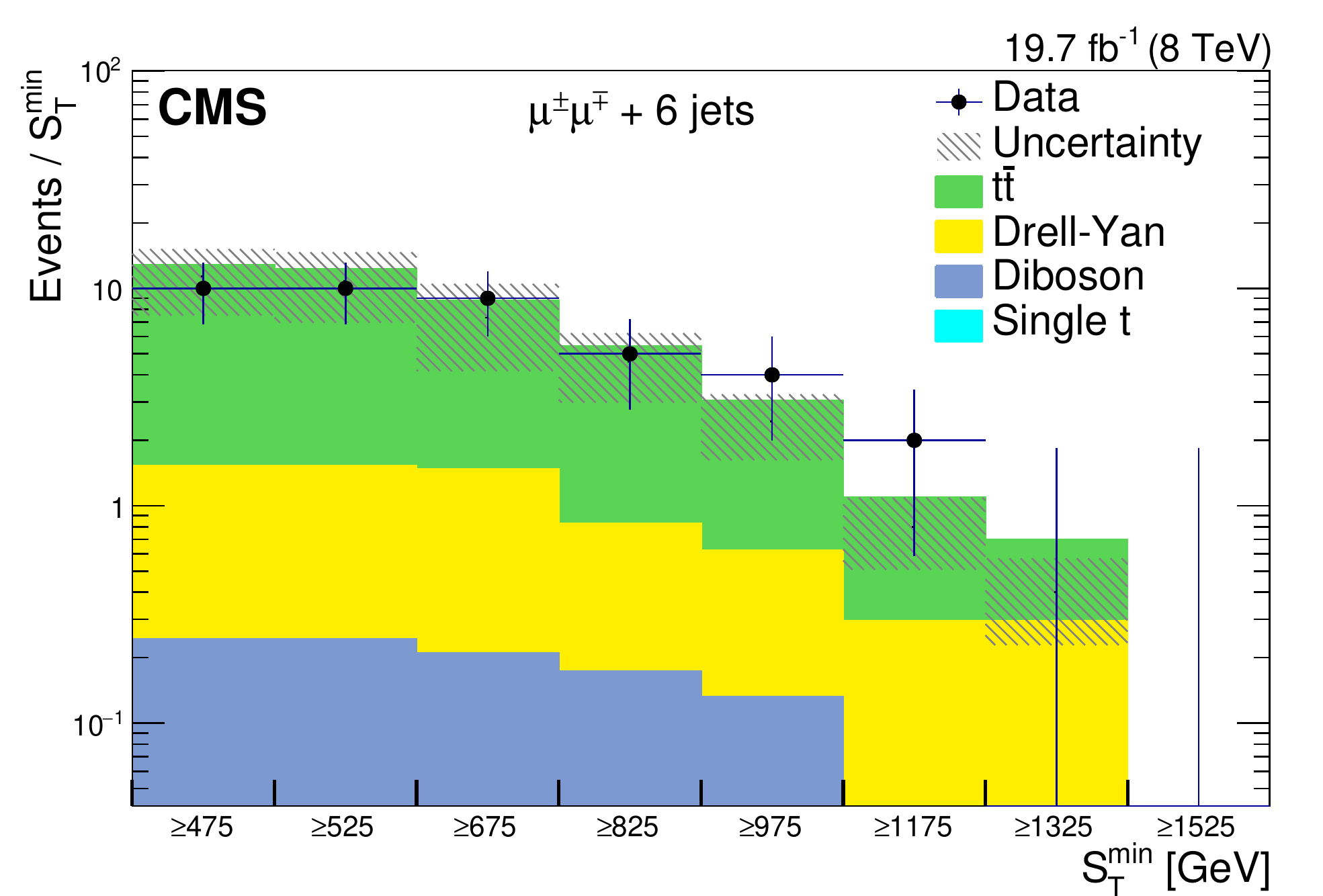}
  \includegraphics[width=0.49\textwidth]{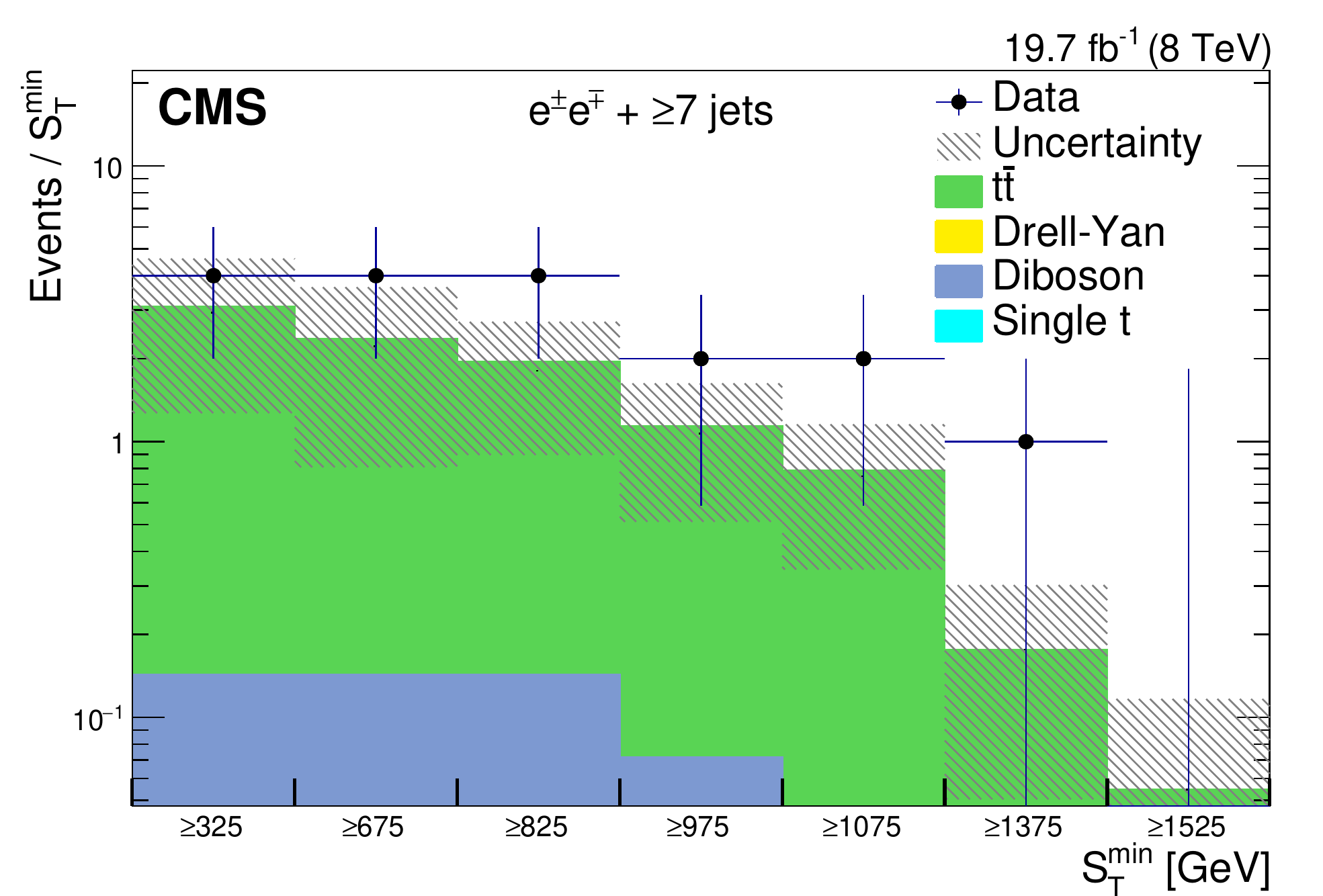} \includegraphics[width=0.49\textwidth]{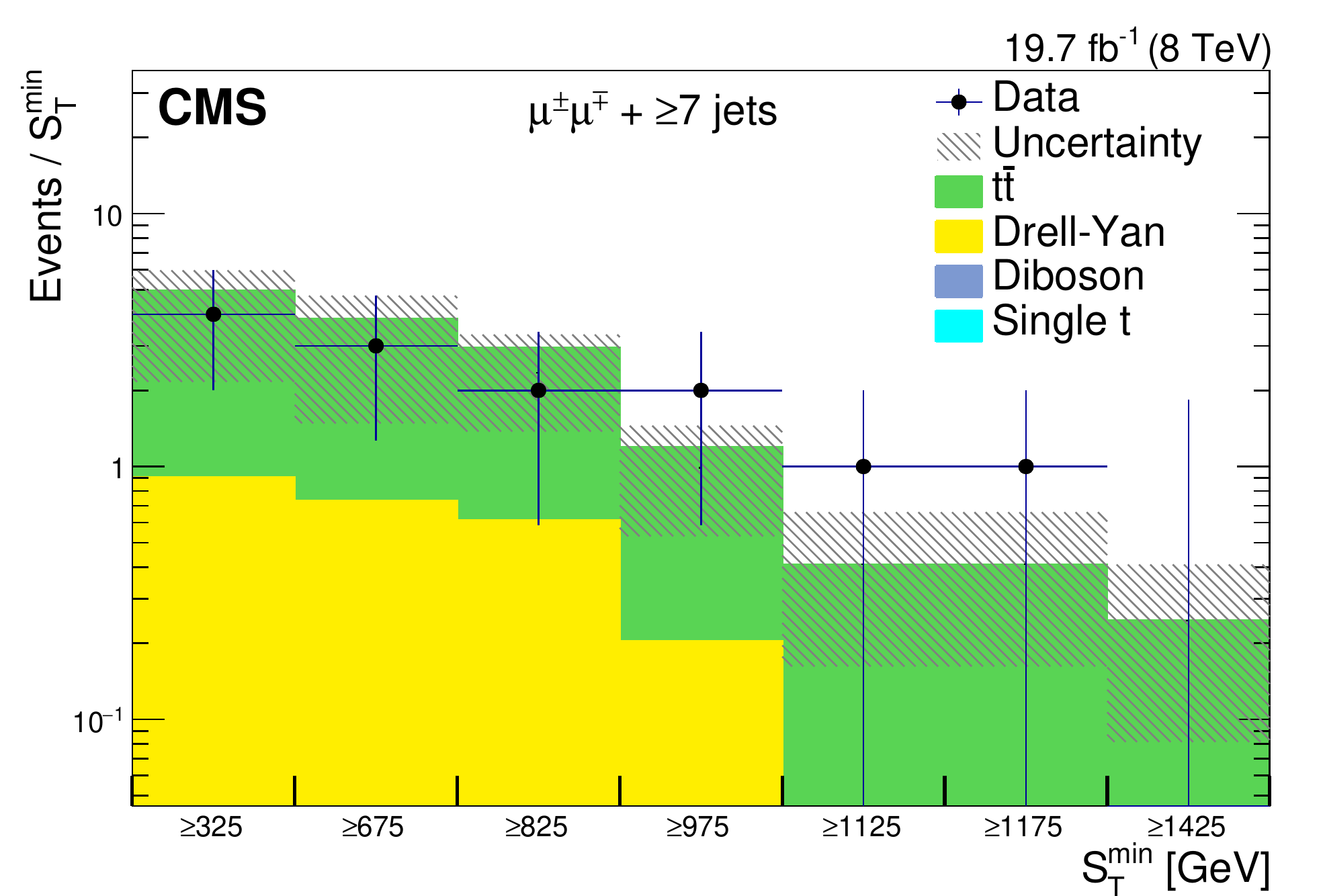}
  \caption{ Events per \stmin for 5th, 6th, and $\geq$7th jets for \ee (left) and \mumu (right). The \stmin selections
    are optimized for \mstop
    ranging from  300 to 1100\GeV. The asymmetric error bars indicate the central confidence intervals for Poisson-distributed data.
  }

\label{fig:stmin}
\end{figure*}

}
\cleardoublepage \section{The CMS Collaboration \label{app:collab}}\begin{sloppypar}\hyphenpenalty=5000\widowpenalty=500\clubpenalty=5000\textbf{Yerevan Physics Institute,  Yerevan,  Armenia}\\*[0pt]
V.~Khachatryan, A.M.~Sirunyan, A.~Tumasyan
\vskip\cmsinstskip
\textbf{Institut f\"{u}r Hochenergiephysik der OeAW,  Wien,  Austria}\\*[0pt]
W.~Adam, E.~Asilar, T.~Bergauer, J.~Brandstetter, E.~Brondolin, M.~Dragicevic, J.~Er\"{o}, M.~Flechl, M.~Friedl, R.~Fr\"{u}hwirth\cmsAuthorMark{1}, V.M.~Ghete, C.~Hartl, N.~H\"{o}rmann, J.~Hrubec, M.~Jeitler\cmsAuthorMark{1}, V.~Kn\"{u}nz, A.~K\"{o}nig, M.~Krammer\cmsAuthorMark{1}, I.~Kr\"{a}tschmer, D.~Liko, T.~Matsushita, I.~Mikulec, D.~Rabady\cmsAuthorMark{2}, B.~Rahbaran, H.~Rohringer, J.~Schieck\cmsAuthorMark{1}, R.~Sch\"{o}fbeck, J.~Strauss, W.~Treberer-Treberspurg, W.~Waltenberger, C.-E.~Wulz\cmsAuthorMark{1}
% ...
\vskip\cmsinstskip
% ...
1:~~Also at Vienna University of Technology, Vienna, Austria\\
2:~~Also at CERN, European Organization for Nuclear Research, Geneva, Switzerland\\
% ...

\end{sloppypar}
\end{document}